\documentclass{PoS}
\usepackage{amsmath,amssymb,mcite}
\usepackage{bbold}

\newcommand{\tr}[0]{\text{tr}}
\newcommand{\ket}[1]{|#1\rangle}
\newcommand{\bra}[1]{\langle#1|}

\newcommand{\path}[1]{[#1]}
\title{Gell-Mann-Oakes-Renner relation for multiple chiral symmetries}

\ShortTitle{Local Ginsparg-Wilson symmetry}

\author{\speaker{Nigel Cundy} and {Weonjong Lee} \\
             Lattice Gauge Theory Research Center, FPRD, and CTP, Department of Physics \&
    Astronomy, Seoul National University, Seoul, 151-747, South Korea\\ 
        E-mail: \email{ndcundy@phya.snu.ac.kr}}

\abstract{As a first step towards establishing a chiral perturbation theory for overlap fermions, we investigate whether there are any ambiguities in the expression for the pion mass resulting from multiple chiral symmetries. The concern is that, calculating the conserved current for Ginsparg Wilson chiral symmetries in the usual way, different expressions of the chiral symmetries lead to different currents. This implies an ambiguity in the definition of the pion and pion decay constant for all Ginsparg-Wilson expressions of the Dirac operator, including the overlap operator. We use a renormalisation group mapping procedure to consider local chiral symmetry transformations for a continuum Ginsparg-Wilson ``Dirac-operator." We find that this naturally leads to an expression for the conserved current that differs from the standard expression by cut-off artefacts, but is independent of which of the Ginsparg-Wilson symmetries is chosen. We recover the standard expressions for the massive Dirac operator, propagator, and chiral condensate. With this in place, we proceed to calculate the pion mass in the mapped theory as a function of the quark mass, and discover a unique expression for $F_\pi$ and $m_\pi$, recovering the usual Gell-Mann-Oakes-Renner relation, baring the substitution of the chiral condensate with its modified value. We hypothesise that the argument can be carried directly over to the lattice theory.}

\FullConference{The XXIVII International Symposium on Lattice Field Theory, Lattice2011\\
		July 10-16, 2011\\
		Squaw Valley, CA, US}

\begin{document}
\section{Introduction}
This project is the start of an attempt to establish an overlap chiral perturbation theory for overlap fermions. First, we need to understand chiral symmetry on the lattice. The challenge is that there is not just one, but an infinite number of chiral symmetries for Ginsparg-Wilson fermions~\cite{Ginsparg:1982bj,*Luscher:1998pqa,*Mandula:2009yd}. These symmetries all agree in the continuum limit, but at non-zero lattice spacing, using standard methods, give non-equivalent conserved currents. This suggests an non-unique definition of the pion, pion decay constant and so on, possibly leading to different measurements of the same quantity on the lattice theory. Although these differences can just be considered lattice artefacts, it is troubling because generally perceived wisdom states that the lattice theory itself is a well defined quantum field theory not only in the continuum limit, and it has been claimed that these ambiguities might imply some serious fundamental problem with Quantum Chromodynamics on the lattice. The goal of this work is to gain a deeper understanding of these issues in the context of the simplest calculation where it might be expected to have an effect: the Gell-Mann-Oakes-Renner relation between the pion mass and the quark mass. We show that the ambiguities arise due to a sub-optimal expression for the conserved current, and that it is possible to formulate a Ginsparg-Wilson theory without this difficulty.    

As we are interested only in the structure of the Ginsparg-Wilson chiral symmetry, the qualitative results of this work should not depend strongly on the choice of the Ginsparg-Wilson Dirac operator. The bulk of this work uses a continuum expression of the Ginsparg-Wilson Dirac operator, $D_R$, as this is the simplest choice using our methods. We will consider the extension to the lattice theory~\cite{Neuberger:1998fp,*Neuberger:1998my,*Narayanan:1993sk,*Narayanan:1993ss} at the end of this work and in more detail in a subsequent work.
\begin{gather}
a D_R^{[a]} = \frac{1}{2}\left(1-e^{i2\arctan(i a D_0)}\right) = \frac{aD_0}{a^2D_0^{\dagger} D_0 + 1} + \frac{a^2D_0^{\dagger} D_0}{a^2D_0^{\dagger} D_0 + 1} = \frac{aD_0}{1+aD_0}
\end{gather}
$D_0$ is the standard Continuum Dirac operator ($D_0\psi = e^{-i g\int dx_\mu A^\mu}\gamma_\nu \partial^\nu (e^{ig \int dx_\mu A^\mu}\psi) $). $a$ is dimensional parameter which we can interpret as an inverse cut-off. $a \rightarrow 0$ corresponds to $D_R \rightarrow D_0$. This operator satisfies the standard Ginsparg-Wilson chiral symmetries.

\section{Ginsparg-Wilson mappings and chiral symmetries}
The notation of this section follows~\cite{Cundy:2009ab,*Cundy:2010pu}.

We have a partition function (for simplicity, we neglect the Yang-Mills term).
\begin{gather}
Z = \int d \overline{\psi}d\psi dU e^{-\overline{\psi} D_0 \psi }.
\end{gather}
We construct a new partition function using the Ginsparg-Wilson mapping procedure
\begin{gather}
Z = \int dU \int d\overline{\psi} d\psi e^{-\overline{\psi}_0 D_0 \psi_0}
\int d \psi_1 d\overline{\psi}_1 e^{(\overline{\psi}_1 - \overline{\psi}_0\hat{\overline{B}})\alpha (\psi_1 - \hat{B} \psi_0)},
\end{gather}
where $\hat{B}$, $\hat{\overline{B}}$ and $\alpha$ are invertible operators acting in the same Hilbert space as the Dirac operators. In a discrete theory, these mappings would therefore represent square matrices: we are not blocking or averaging to, for example, reduce from the continuum to the lattice, but constructing a different expression of the Dirac operator in the same space-time.
These mappings are integral transformations of the fields
\begin{gather}
(\hat{B}\psi)_a(x) \equiv \int d^4x' \hat{B}_{ab}(x,x') \psi_b(x'),
\end{gather}
for coordinates $x$ and spinor indices $a$ and $b$. Throughout this work, I will only need to consider invertible mappings, where the inverse is defined as 
\begin{gather}
(B\psi)_b(x) \equiv \int d^4x' {B}_{ab}(x,x') \psi_b(x')
\end{gather}
The kernel ${B}_{ab}(x,x')$ satisfies
\begin{gather}
\int d^4x' {B}_{ab}(x,x')\hat{B}_{bc}(x',x'') = \delta^{(4)}(x-x'') \delta_{ac},
\end{gather}
and we subsequently write $\hat{B}$ as $B^{-1}$. These mappings are functions of the gauge fields and contain a non-trivial spinor structure. We then integrate over the fields $\psi_0$ to give a new Lagrangian, $\mathcal{L}_1 = \overline{\psi}_1D_1\psi_1 + \tr\log[B \overline{B}]$, where we absorb $\tr\log[B \overline{B}]$ into the gauge action. 
In practice, there will be a family of mapping operators which generate the same Dirac operator $D_1$, which we parametrise as $B^{(\eta)}$, $\overline{B}^{(\eta)}$ and $\alpha^{(\eta)}$. We will consider those blockings where $\alpha \rightarrow \infty \mathbb{1}$. The mapped theory will then obey a Ginsparg-Wilson chiral symmetry defined by
\begin{align}
0= &\gamma_L^{(\eta)} D_1 + D_1\gamma_R^{(\eta)} & D_1 =& \overline{B}^{(\eta)} D_0 B^{(\eta)}\nonumber\\
B^{(\eta)} =& D_0^{-(\eta + 1)/2} D_1^{(\eta + 1)/2}&\overline{B}^{(\eta)} = & D_1^{(1-\eta)/2} D_0^{-(1-\eta)/2}.\nonumber\\
\gamma_L^{(\eta)} = &\overline{B}^{(\eta)}\gamma_5(\overline{B}^{(\eta)})^{-1} =(1-2D)^{\frac{\eta - 1}{2}} \gamma_5&
\gamma_R^{(\eta)} = &   (B^{(\eta)})^{-1}\gamma_5 B^{(\eta)} = \gamma_5 (1-2D)^{\frac{\eta + 1}{2}},
\end{align}
Locality of $\gamma_L^{(\eta)}$ and $\gamma_R^{(\eta)}$ requires an odd integer value of the parameter $\eta$. The choice of the chiral symmetry is equivalent to the choice of $B$ and $\overline{B}$ and therefore the choice of $\eta$.

The mappings $B$ and $\overline{B}$ are analytic around $p_\mu = 0$, the only zero in $D_0$ and $D_1$. There are difficulties for UV momentum, i.e. infinite eigenvalues of $D_0$ or eigenvalues 1 of $D_1$, caused ultimately because we are mapping between a regulated theory and an unregulated unrenormalised continuum theory: we expect divergences in the UV. In this work, we shall neglect these difficulties for simplicity. In a complete calculation, which we will give in a fuller paper, this issue is resolved by using a regulated Dirac operator, $D_R^{[a_\epsilon]}$, as the basis for the mapping rather than $D_0$. This requires using the Ginsparg-Wilson chiral symmetry for $D_R^{[a_\epsilon]}$ in the derivation of the Ginsparg-Wilson chiral symmetry for $D_1$. Demanding self-consistency, the correct limit as $a \rightarrow a_\epsilon$ and independence of the inverse cut-off $a_\epsilon$ restricts the Ginsparg-Wilson chiral symmetry to the expressions derived here from a naive use of the continuum operator. 

\section{Propagator renormalisation}
2-point functions can be calculated from the generating functional
\begin{align}
\langle {\psi}(x) \overline\psi(y) \rangle =& \lim_{J\rightarrow 0}\frac{\partial^2}{\partial \overline{J}(x) \partial {J} (y)} \log \int d U d\overline{\psi}_0d\psi_0  e^{\overline{\psi}_0 D_0 \psi_0 + \overline{J} \psi_0 + \overline{\psi}_0 J}\nonumber\\
=& \lim_{J\rightarrow 0}\frac{\partial^2}{\partial J(y) \partial \overline{J} (x)} \log \int d U d\overline{\psi}_1^{(\eta)}d\psi_1^{(\eta)}  e^{\overline{\psi}_1^{(\eta)} D_1 \psi_1^{(\eta)} + \overline{J} B^{(\eta)} \psi_1^{(\eta)} + \overline{\psi}_1^{(\eta)} \overline{B}^{(\eta)} J}.
\end{align}
Therefore, the propagator, $S$ in the mapped theory is 
\begin{gather}
S =  B^{(\eta)}D_1^{-1} \overline{B}^{(\eta)}.
\end{gather}
For example, if $D_1 = D_R$, then $D_0 = D_R/(1-aD_R)$ and $\overline{B}^{(\eta)} B^{(\eta)} = D_R/D_0 = 1-aD_R$. This leads to the standard O($a$) improvement factor, $1-aD$, required to maintain the chiral symmetry of the Ginsparg-Wilson propagator~\cite{Capitani:1999ay,*Chiu:1998eu}.

\section{Conserved Current}

The conserved current in the continuum theory is found by applying an infinitesimal local chiral symmetry transformation,
\begin{align}
\overline{\psi}_0 \rightarrow&  \overline{\psi}_0(1+i\epsilon(x) \gamma_5) & \psi_0 \rightarrow (1+i\epsilon(x) \gamma_5) \psi_0
\end{align}
which gives
\begin{gather}
J^\mu(x) = \frac{\partial}{\partial\partial_\mu \epsilon(x)}(\overline{\psi}(1+i\epsilon(x) \gamma_5) D_0 (1+i\epsilon(x) \gamma_5) \psi
\end{gather}
or
\begin{gather}
J^\mu(x) = \overline{\psi}(x) \gamma_5 \gamma^\mu \psi(x).
\end{gather}
In the mapped theory, the chiral symmetry transformation is constructed by applying $\psi_0^{(\eta)} = B^{(\eta)}\psi_1^{(\eta)}$ and $\overline{\psi}_0^{(\eta)} = \overline{\psi}_1^{(\eta)}\overline{B}^{(\eta)}$:
\begin{align}
\overline{\psi}_1^{(\eta)} \rightarrow&  \overline{\psi}_1^{(\eta)}(1+i\overline{B}^{(\eta)}\epsilon(x)(\overline{B}^{(\eta)})^{-1} \overline{B}^{(\eta)}\gamma_5(\overline{B}^{(\eta)})^{-1}) =\overline{\psi}_1^{(\eta)} (1+i\epsilon_L^{(\eta)} \gamma_L^{(\eta)})
\nonumber\\
 \psi_1^{(\eta)} \rightarrow& (1+i(B^{(\eta)})^{-1}\epsilon(x)B^{(\eta)} (B^{(\eta)})^{-1}\gamma_5B^{(\eta)}) \psi_1^{(\eta)} = (1+i \epsilon_R^{(\eta)} \gamma_R^{(\eta)}) \psi_1^{(\eta)}
\end{align}
Note that $\epsilon_R$ and $\epsilon_L$ are now non-degenerate: this is the principle difference between this and previous approaches to construct the current in Ginsparg-Wilson theories. $\epsilon_R^{(\eta)}$ for different $\eta$s differ by O($a$) effects. The choice of $\epsilon_R^{(\eta)}$ is fixed for each choice of chiral symmetry.
It is now an easy matter to calculate the conserved current,
\begin{align}
J_\mu(x) =& \frac{\partial}{\partial \partial^\mu \epsilon(x)}\left(\overline{\psi}^{(\eta)} (\gamma_L^{(\eta)} \epsilon_L^{(\eta)} D_1 + D_1 \epsilon_R^{(\eta)}\gamma_R^{(\eta)})\psi^{(\eta)}\right) \nonumber\\
=&\frac{\partial}{\partial \partial^\mu \epsilon(x)}\left(\overline{\psi}^{(\eta)} ({B}^{(\eta)})^{-1} B^{(1)}\overline{B}^{(1)} \gamma_\mu \partial_\mu \epsilon \gamma_5 B^{(\eta)}  \psi^{(\eta)}\right).
\end{align}
Given that $\psi^{(\eta_2)} = ({B}^{(\eta_2)})^{-1}({B}^{(\eta_1)})\psi^{(\eta_1)}$ and $\overline{\psi}^{(\eta_2)} = \overline{\psi}^{(\eta_1)} (\overline{B}^{(\eta_1)})^{-1}(\overline{B}^{(\eta_2)})$, this expression is independent of $\eta$, and which of the chiral symmetries is used. The expectation value for the current then becomes the simple expression,
\begin{gather}
\langle J_\mu(x)\rangle = \left\langle\frac{\partial}{\partial \partial^\mu \epsilon(x)} \tr \frac{1-D}{D} \gamma_\mu \gamma_5 \partial_\mu \epsilon \right\rangle.
\end{gather}
 The standard $\eta$-dependant expression~\cite{Kikukawa:1998py} is derived from $\frac{\partial}{\partial (\partial_\mu \epsilon)}\overline{\psi}^{(\eta)}_1 (\gamma_L^{(\eta)} {\epsilon_L^{(1)}} D_1 + D_1 {\epsilon_R^{(-1)}}\gamma_R^{(\eta)})\psi^{(\eta)}_1$. In the context of the mapping formulation, this is unnatural as it mixes different $\eta$s, although it only differs from the expression recommended here by O($a$) artefacts.

\section{Gell-Mann Oakes Renner relation}

%
The current is the generator of the pion field
 \begin{gather}
 \bra{0} J^\mu(x) \ket{\pi} = i \frac{F_\pi p^\mu e^{ipx}}{(2\pi)^{3/2} \sqrt{2p^0}}.
 \end{gather}
Suppose that the symmetry breaking term in the action is
\begin{gather}
H_1 = \sum_n u_n \Phi_n,
\end{gather}
where $\Phi$ is some field which transforms under the symmetry as
$
[T^a,\Phi_n] = - \tau^a_{nm}\Phi_m,
$
 $u_n$ is a coefficient, and $\tau$ is the generator of an appropriate representation of the chiral symmetry.
The mass matrix for the Goldstone Bosons of a spontaneously broken broken symmetry is~\cite{Weinburg}
\begin{gather}
M_{cd}^2 = - F^{-1}_{ca} F^{-1}_{db} \langle [T^a,[T^b,H_1]]\rangle_0,
\end{gather}
with $F_{ab} = F_\pi \delta_{ab}$. The fermionic Lagrangian for the Ginsparg-Wilson mapped theory for two quark flavours is
\begin{gather}
\mathcal{L}_f = \overline{\psi}^{(\eta)}_u[ D_1 +  \overline{B}^{(\eta)} B^{(\eta)}\mu_u]\psi^{(\eta)}_u + \overline{\psi}^{(\eta)}_d[D_1 + \overline{B}^{(\eta)} B^{(\eta)} \mu_d]\psi^{(\eta)}_d \label{eq:Lag}
\end{gather}
With $\overline{B} B \rightarrow (1-D)$ this gives the conventional form for the massive Ginsparg-Wilson Dirac operator: $D_1[\mu] = D_1 + \mu (1-D_1)$.
The symmetry breaking terms in the Lagrangian can be written as
\begin{gather}
H_1 = \Phi^4 (\mu_u + \mu_d) + \overline{\Phi}^3(\mu_u - \mu_d),\label{eq:symb}
\end{gather}
where the chiral 4 vectors are are given in terms of the iso-spin Pauli matrices $t^a = \sigma^a/2$
\begin{align}
\Phi^a =& i \overline{q}\hat{\gamma}_5 t^a q&\Phi^4 = & \frac{1}{2} \overline{q}\hat{1}q\nonumber\\
\overline{\Phi}^a = & \overline{q} t^a \hat{1}q& \overline{\Phi}^4 =&- i\frac{1}{2} \overline{q} \hat{\gamma}_5 q & 
q = \left(\begin{array}{c}\psi_u\\\psi_d\end{array}\right),
\end{align}
and $\hat{\gamma}_5$ and $\hat{1}$ are operators which commute with $t^a$. In the standard theory, these are just $\gamma_5$ and the identity operator. In the mapped theory, a natural possibility is to use
\begin{align}
\hat{\gamma}_5^{(\eta)} = & \overline{B}^{(\eta)} {\gamma}_5 B^{(\eta)}& \hat{1} =& \overline{B}^{(\eta)} B^{(\eta)},
\end{align}
so the symmetry breaking term in the Lagrangian (\ref{eq:Lag}) agrees with equation (\ref{eq:symb}). This gives
\begin{align}
\hat{\gamma}_5^{(\eta)}\gamma_R^{(\eta)} =& \gamma_L^{(\eta)} \hat{\gamma}_5^{(\eta)} = \hat{1}\nonumber\\
\hat{1}\gamma_R^{(\eta)} =& \gamma_L^{(\eta)} \hat{1} = \hat{\gamma}_5^{(\eta)},
\end{align}
and the modified chiral vectors satisfy the usual continuum transformation law
\begin{align}
[X^a,\Phi^b] =&-\delta^{ab}\Phi^4&
[X^a,\overline{\Phi}^b] =& - \delta^{ab} \overline{\Phi}^4\nonumber\\
[X^a,\Phi^4]=&\Phi^a&
[X^a,\overline{\Phi}^4]=& \overline{\Phi}^a.
\end{align}
We then can use a standard calculation to recover
\begin{gather}
m_\pi^2 = - 4 (\mu_u + \mu_d) \langle \Phi^4\rangle_0/F_\pi^2.
\end{gather}
Therefore, if the chiral four vectors and currents are local and well defined, the GMOR relation is satisfied in the mapped theory. We only need to modify the definition of $\langle \Phi^4\rangle$ and $F_\pi$.

If $D_1$ is the regulated continuum operator $D_R$, we obtain the $\eta$-independent expressions
\begin{align}
\Phi^a = & i \overline{q}_1^{(\eta)} (1-D_R) \gamma_5 (1-2D_R)^{\frac{1+\eta}{2}} t^a q_1^{(\eta)} 
&
\Phi^4 = & \frac{1}{2} \overline{q}_1^{(\eta)} (1-D_R) q_1^{(\eta)}
\nonumber\\
\overline{\Phi}^a = & \overline{q}^{(\eta)}_1 t^a (1-D_R) q^{(\eta)}
&
\overline{\Phi}^4 = &-i \frac{1}{2} \overline{q}_1^{(\eta)}(1-D_R) \gamma_5 (1-2D_R)^{\frac{1+\eta}{2} }q^{(\eta)}_1
\end{align}
The chiral condensate, extracted from $\Phi^4$, is in agreement with previous results~\cite{Hasenfratz:1998jp,*Neuberger:1997bg}.
\section{Application to the lattice overlap operator}
Clearly, carrying this framework to the lattice overlap operator contains a number of complications. Firstly, the lattice overlap operator does not commute with the continuum overlap operator; secondly the naive rectangular blockings from the continuum to the lattice are not invertible; thirdly, we need to ensure that we correctly regulate the $1/(1-D)$ term to avoid complications with the $D=1$ eigenvalues. Of these, the second complication is at first glance the largest challenge. It can be resolved using a two step mapping/blocking procedure. In the first step, a continuum Dirac operator is constructed which has the lattice operator embedded within it. Space time is decomposed into regions around the lattice sites $L$ and the bulk $B$. The Dirac operator is constructed so that the $LL$ interactions resemble that of the lattice theory, while there will also be $BB$ and $LB$ interactions. As $D_1$ and $D_0$ are in the same Hilbert space, a mapping rather than blocking or averaging procedure is required to construct $D_1$ from $D_0$, and there is no obvious reason to suppose that there is not some suitable choice of $D_1$ which allows invertible mappings. One possibility is explored in ~\cite{Cundy:2009ab,*Cundy:2010pu}. The construction of the chiral symmetry, chiral condensate, and currents then proceeds as outlined above. A second blocking transformation is used to extract the lattice operator from $D_1$. If this blocking transformation (which we may define in terms of operators $R$ and $\overline{R}$, so that $\psi_2 = R \psi_1$ and $\overline{\psi}_2 = \overline{\psi_1} \overline{R}$) commutes with $\gamma_5$ and the eigenvalues of $\overline{R}R$ are one (corresponding to the lattice sites) or zero (corresponding to the bulk), this will leave the form of the chiral symmetry intact and all the conclusions of this work unchanged.

We therefore expect that the results of this work may also be extended to the lattice theory (except for modifications to to additional dependence on $D^{\dagger} D$ in the mapping). In particular, the conserved current and chiral four vectors $\Phi$ and $\overline{\Phi}$ will be independent of the choice of the Ginsparg-Wilson chiral symmetry given that the fermion fields are also dependent on $\eta$. If this holds, then there would be no ambiguity when the lattice theory is correctly formulated, and no reason to suspect that there is a disease in the lattice formulation caused by the infinite group of chiral symmetries.
\section{Conclusions}
By using a RG mapping construction, we have reconstructed the standard form for the massive Dirac operator, chiral condensate and quark propagator for a Ginsparg-Wilson Dirac operator, and shown that the same observables are obtained regardless of the choice of chiral symmetry (expressed in terms of choosing the parameter $\eta$). The standard conserved current is not independent of $\eta$, however we have shown that the local current is independent of $\eta$ and is derived naturally from this procedure. With this machinery in place, it is straight forward to repeat the continuum derivation of the Gell-Mann-Oakes-Renner relation. This leads to a unique lattice definition of $\pi$, $m_\pi$, $f_\pi$ etc., and we do not see that the multiple Ginsparg-Wilson will have any effect on the physics of the theory even an non-zero inverse cut-off. Although our construction used a particular continuum Dirac operator, we expect that the results will carry over to any Ginsparg-Wilson Dirac operator, including the lattice overlap operator, as will be discussed in a forthcoming paper. 
   
\section*{Acknowledgements}
Funding was provided by the BK21 program of the NRF, Republic of Korea. The research of W. Lee is supported by the Creative Research Initiatives program (3348-20090015) of the NRF grant funded by the Korean government (MEST). W. Lee would like to acknowledge the support from KISTI supercomputing center through the strategic support program for the supercomputing application research [No. KSC-2011-C3-03].

\bibliographystyle{elsarticle-num-mcite}
\bibliography{weyl}

\end{document}